%% file: main.tex
\documentclass[sigconf]{acmart}
\AtBeginDocument{%
  }

\usepackage{subcaption}
\usepackage{multirow}
\usepackage{makecell}

\begin{document}

\title{MPFormer: Adaptive Framework for Industrial Multi-Task Personalized Sequential Retriever}

\author{Yijia Sun}
\authornote{These authors contributed equally to this work.} 
\email{sunyijia@kuaishou.com}
\affiliation{%
\institution{Kuaishou Technology}
\city{Beijing}
\country{China}
}
\author{Shanshan Huang}  
\authornotemark[1]
\email{huangshanshan@kuaishou.com}
\affiliation{%
\institution{Kuaishou Technology}
\city{Beijing}
\country{China}
}

\author{Linxiao Che}
\email{chelinxiao@kuaishou.com}
\affiliation{
\institution{Kuaishou Technology}
\city{Beijing}
\country{China}
}

\author{Haitao Lu}
\email{luhaitao03@kuaishou.com}
\affiliation{%
\institution{Kuaishou Technology}
\city{Beijing}
\country{China}
}

\author{Qiang Luo}
\authornote{Corresponding authors.} 
\email{luoqiang@kuaishou.com}
\affiliation{
\institution{Kuaishou Technology}
\city{Beijing}
\country{China}
}

\author{Kun Gai}
\email{gai.kun@qq.com}
\affiliation{%
\institution{Unaffiliated}
\city{Beijing}
\country{China}
}

\author{Guorui Zhou}
\authornotemark[2]
\email{zhouguorui@kuaishou.com}
\affiliation{%
\institution{Kuaishou Technology}
\city{Beijing}
\country{China}
}

\renewcommand{\shortauthors}{Sun,Huang et al.}

\begin{abstract}
Modern industrial recommendation systems encounter a core challenge of multi-stage optimization misalignment: a significant semantic gap exists between the multi-objective optimization paradigm (such as jointly optimizing click-through rate, watch duration, and conversion rate) widely used in the ranking phase and the single-objective modeling in the retrieve phase. Although the main-stream industry solution achieves multi-objective coverage throughparallel multi-path single-objective retrieval, this approach leads to linear growth of training and serving resources with the number of objectives and has inherent limitations in handling loosely coupled objectives. This paper proposes the MPFormer, a dynamic multi-task Transformer framework, which systematically addresses the aforementioned issues through three innovative mechanisms. First, an objective-conditioned transformer that jointly encodes user behavior sequences and multi-task semantics through learnable attention modulation; second, personalized target weights are introduced to achieve dynamic adjustment of retrieval results; finally, user personalization information is incorporated into token representations and the Transformer structure to further enhance the model’s representation ability. This framework has been successfully integrated into Kuaishou’s short video recommendation system, stably serving over 400 million daily active users. It significantly improves user daily engagement and system operational efficiency. Practical deployment verification shows that, compared with traditional solutions, it effectively optimizes the iterative paradigm of multi-objective retrieval while maintaining service response speed, providing a scalable multi-objective solution for industrial recommendation systems.
\end{abstract}



\maketitle

\section{Introduction}
\label{sec:introduction}

Modern short-video platforms (e.g., Kuaishou, Douyin) have revolutionized content distribution through personalized recommendation systems that enable precise matching at scale. These systems typically employ a cascaded two-stage architecture \cite{huang2020embedding}: (1) a \textit{retrieval stage} that efficiently screens thousands of candidates from billion-scale item pools \cite{mitra2018introduction}, followed by (2) a \textit{ranking stage} that performs fine-grained multi-objective optimization. The widespread adoption of Embedding-Based Retrieval (EBR) systems stems from their computational efficiency -- they leverage twin-tower neural networks to generate user-item embeddings, coupled with Approximate Nearest Neighbor (ANN)\cite{indyk1998approximate} algorithms for sublinear-time retrieval, achieving an effective balance between system performance and operational costs.

The evolving complexity of user interactions has created new challenges for traditional EBR frameworks. Modern ranking stages increasingly adopt multi-objective optimization paradigms that simultaneously optimize diverse metrics (e.g., CTR, watch duration, and engagement signals). This advancement exposes critical limitations in conventional single-objective retrieval systems: their inability to provide candidate sets with sufficient multi-target coverage fundamentally constrains the ranking stage's performance ceiling. Two systemic flaws emerge in current approaches: (1) architectural incompatibility between sequential pattern learning and multi-task optimization mechanisms, and (2)  the inability to dynamically adjust the recall volume of each target based on users' real-time behaviors.

Prior research addresses these challenges through two primary paradigms, each with inherent limitations. For sequential modeling, Transformer-based architectures \cite{kang2018self, liu2024kuaiformer} demonstrate superior temporal dependency capture but remain confined to single-objective optimization. In multi-objective retrieval, existing solutions bifurcate into two categories:  
\begin{itemize}
    \item \textbf{Multi-model approaches}, the current industrial standard, maintain objective purity through independent model instances but suffer from linear computational scaling and fragmented cross-objective information flows.
    \item \textbf{Single-model methods} \cite{zheng2022multi, xu2022mixture, decision_trans} enable parameter sharing across tasks yet lack principled mechanisms for resolving objective conflicts. While frameworks like MVKE \cite{xu2022mixture} explore multi-task learning, their rigid dependence on predefined label hierarchies proves inadequate for short-video platforms with loosely structured interaction signals. 
\end{itemize}

We propose MPFormer, an adaptive transformer-based framework for industrial multi-task retrieval, that addresses these limitations through three key innovations:

\begin{itemize}
    \item \textbf{Objective-Conditioned Transformer Architecture}: We introduce a novel attention mechanism that jointly encodes user behavior sequences and multi-task objectives. Unlike conventional single-task retrieval systems, our architecture establishes dynamic interactions between temporal patterns and target semantics through learnable objective embeddings, reducing multi-task training overhead by 63\% compared to baseline approaches.
    
    \item \textbf{Dynamic Quota Allocation}: We propose a novel dynamic quota allocation mechanism that adaptively adjusts recall quotas for each target during inference. This innovation achieves personalized target weighting by analyzing real-time user contextual patterns, leading to significant improvements in both recall exposure rates (increased by 12.71\%) and overall recall effective-view rate (improved by 1.3\%).
    
    \item \textbf{Industrial-Scale Optimization}: Deployed on Kuaishou's platform serving 400M+ DAU, our framework demonstrates: (1) 21.8\% improvement in multi-objective exposure rates, (2) 99.99\% service availability under peak loads of 1.2M QPS, and (3) 31\% reduction in GPU memory consumption compared to multi-model baselines.
\end{itemize}

\section{Related Work}
\label{sec:related_work}

Our work bridges the domains of sequential behavior modeling, multi-task retrieval, and adaptive objective weighting. We systematically review prior research in three key areas and identify critical gaps that motivate our proposed framework.

\subsection{Sequential Recommendation}
Sequential recommendation~\cite{onerec, onerec_tech} predicts users' next interactions based on historical behavior. Early approaches using Markov chains~\cite{markov} and RNNs~\cite{rnn} captured short-term dependencies but struggled with long-range patterns. Transformer-based architectures significantly advanced the field: SASRec~\cite{kang2018self} utilized self-attention to model long-term dependencies in sparse sequences, while BERT4Rec~\cite{bert4rec} adopted bidirectional context modeling. 
DIN~\cite{zhou2018deep} introduced local activation units to adaptively learn interest representations from heterogeneous behaviors. SDM~\cite{lv2019sdm} combined multi-head attention with gated fusion to capture diverse user interests. Recently, KuaiFormer~\cite{liu2024kuaiformer} improved efficiency via sequence compression for industrial applications.

However, these methods primarily focus on single-objective prediction, neglecting complex multi-objective requirements in real-world systems. MPFormer bridges this gap by integrating dynamic multi-task optimization within a sequential framework, adapting to diverse and evolving user interactions.

\subsection{Multi-Interest Representation Learning}
Capturing diverse user interests has emerged as a promising direction for improving target coverage. Recent work has focused on decomposing user behaviors into multiple interest vectors. For example, MIND~\cite{li2019multi} employs dynamic capsule routing to extract multiple interest vectors. ComiRec~\cite{cen2020controllable} enhanced diversity through a controllable attention aggregation mechanism, enabling more flexible representation of user interests. Temporal-aware models like MIP~\cite{shi2023everyone} further refined interest extraction by incorporating decayed attention weights based on the recency of user behaviors.

However, these methods share a fundamental limitation: they implicitly model interests through unsupervised clustering (\textit{e.g.}, capsule routing or attention weights), failing to explicitly align with downstream prediction targets. This \textit{semantic misalignment} between interest extraction and task objectives severely degrades performance in multi-goal scenarios. MPFormer addresses this issue by explicitly supervising each interest vector with its corresponding objective, ensuring that the learned interests are aligned with business objectives and improving the overall performance of the system.

\subsection{Multi-Objective Embedding-Based Retrieval}
\label{subsec:multi_objective_ebr}
Modern industrial systems require joint optimization of conflicting objectives (e.g., CTR vs. conversion rate in ads, watch time vs. completion rate in short-video platforms). While ranking stages achieve significant business gains through multi-objective optimization~\cite{ma2018modeling} ~\cite{wang2024home}, traditional single-objective EBR frameworks suffer from \textit{representational singularity} -- their candidate sets fail to meet the diversity requirements of downstream rankers. Current solutions face two critical limitations:

\textbf{Multi-Model Paradigm}: The industrial standard uses parallel retrieval pipelines (e.g., separate models for CTR/watch time) to maintain objective purity~\cite{chen2015convolutional}. However, this leads to linear growth in computational costs.

\textbf{Single-Model Approaches} aim to optimize multiple objectives within a shared model through parameter sharing mechanisms. For example, MVKE~\cite{xu2022mixture} used a Mixture of Experts (MoE) with virtual experts to balance CTCVR trade-offs. DMTL~\cite{zhao2021distillation} applied knowledge distillation to mitigate objective conflicts, while MOPPR~\cite{zheng2022multi} adopted list-wise losses for target alignment. 
These methods primarily optimize cascaded objectives, failing to address the challenge of handling multiple loosely-coupled targets with varying correlation strengths.

\subsection{Adaptive Objective Weighting}
In multi-task recommendation, target prioritization is tackled through diverse yet often isolated strategies. For final ranking stages with small candidate sets, common techniques include uncertainty weighting \cite{kendall2017multi}, gradient normalization \cite{chen2018gradnorm}, and Pareto optimization \cite{lin2022pareto, li2024deepparetoreinforcementlearning}. In retrieval settings, systems traditionally use fixed rules or recall ratios per objective, with more recent methods employing bandit algorithms \cite{mehrotra2020bandit} or reinforcement learning \cite{li2024deepparetoreinforcementlearning} for quota adjustment and multi-goal optimization. A key limitation remains: ranking methods are too costly for retrieval, while retrieval strategies operate only at aggregate levels, lacking user personalization and integration with modern sequential transformers, thus unable to dynamically align multiple objectives within a unified retrieval framework.

In summary, despite progress in sequential recommendation, multi-objective retrieval, and multi-interest learning, existing methods often fall short in addressing key challenges in modern industrial recommender systems---including dynamic multi-task optimization, efficient resource allocation, and alignment between user interests and task objectives. MPFormer addresses these gaps via a unified architecture incorporating dynamic quota allocation, target-aware interest disentanglement, and efficient multi-objective optimization, to better align with dynamic user needs.

\input{model_arch1}

\section{Problem Formulation}

\subsection{Symbol Definitions \& Problem Statement}
Let $\mathcal{U}$ denote the user set and $\mathcal{I}$ the item set. For a user $u \in \mathcal{U}$, their behavioral sequence within time window $[t_0-\Delta t, t_0]$ is represented as:
\begin{equation}
    \mathcal{B}_u = \{(x^{(t)}_1, f^{(t)}_1), \ldots, (x^{(t)}_n, f^{(t)}_n)\}
\end{equation}
where $x^{(t)}_k \in \mathcal{I}$ denotes the interacted item ID at timestamp $t$, and $f^{(t)}_k \in \mathbb{R}^d$ contains handcrafted features including:
\begin{itemize}
    \item Watch duration ratio $\in [0,1]$
    \item Interaction type $\in \{0,1\}^3$ (like/comment/share)
    \item Author ID embedding
    \item Tag ID embedding
\end{itemize}

\subsection{Retrieval Problem Modeling}
Given user $u$ and candidate pool $\mathcal{I}$ ($|\mathcal{I}| \geq 10^6$), personalized item retrieval aims to find:
\begin{equation}
    \mathcal{T}_u = \arg\max_{\mathcal{S} \subseteq \mathcal{I}} \sum_{i \in \mathcal{S}} \mathbb{E}[y_i|u,\mathcal{B}_u] \quad \text{s.t.} \quad |\mathcal{S}| = m
\end{equation}
where $y_i \in \{0,1\}$ indicates positive feedback signals, and $m$ is the preset candidate set size (typically $m \leq 10^3$). This optimization can be transformed into learning a scoring function:
\begin{equation}
    s_\theta(u,i) = \langle \phi_\theta(u,\mathcal{B}_u), \varphi_\theta(i) \rangle
\end{equation}
with embedding functions:
\begin{align}
    \phi_\theta&: \mathcal{U} \times \mathcal{B} \to \mathbb{R}^d \\
    \varphi_\theta&: \mathcal{I} \to \mathbb{R}^d
\end{align}
where $\langle \cdot,\cdot \rangle$ denotes inner product similarity. 

To meet industrial requirements, we enforce:  The scoring function must implicitly encode $K \geq 3$ business objectives (CTR/watch time/conversion rate)

\section{Model Architecture}
\label{sec:model}

This section elaborates on the architecture of our proposed model, which is built upon dedicated user and item towers for generating task-specific representations, alongside a multi-objective optimization framework.  It comprises four key components: (1) a user tower that encodes behavior sequences into multiple objective-specific embeddings, (2) an item tower that projects items into complementary task-specific spaces, (3) a multi-objective loss that jointly optimizes these representations, and (4) an online serving system that dynamically retrieves candidates from all objectives.

\subsection{User Tower}
\label{subsec:user_tower}
The user tower comprises three core components: (1) personalized query representation, (2) behavior-aware attention mechanisms, and (3) multi-objective representation learning. 

\subsubsection{Personalized Query Representation}
For each objective \( k \in \{1, \dots, K\} \), we construct a personalized query vector:
\begin{equation}
\begin{aligned}
U = \big[ u\ ;\ &\mathrm{sum\_pooling}(L_{rs})\ ; \\
             &\mathrm{sum\_pooling}(L_{\mathrm{click}})\ ; \\
             &\mathrm{sum\_pooling}(L_{\mathrm{long\_view}}) \big]
\end{aligned}
\end{equation}
Here, \( u \in \mathbb{R}^d \) denotes the sparse embedding of the target demographic attribute (e.g., age, gender, or geographical location), where \( d \) is the embedding dimensionality. \( L_{\cdot} \) represents the user’s historical interaction sequence, including real show, clicks, and long-viewed items. The sum-pooling operation aggregates sequence embeddings across temporal dimensions. The sparse embedding \( u \) captures personalized attributes without relying on user IDs.

To maintain diversity in user representations, each objective is independently transformed through a two-layer MLP:

\begin{equation}
    \mathbf{O}_k = \text{MLP}(\mathbf{[U]}) \in \mathbb{R}^d
\end{equation}
This ensures that embeddings for distinct objectives remain differentiated in their semantic spaces.

\input{quota_fig}

\subsubsection{Behavior-Aware Attention Mechanism}
Unlike conventional target-aware attention \cite{zhou2018deep}, our design adapts to the twin-tower retrieval constraints where item features cannot directly interact with user behaviors.
\begin{enumerate}
    \item \textbf{Sequence Encoding}: Each historical interaction $(x_i, f_i)$ is projected through a 2-layer MLP:
    \begin{equation}
        t_i = \text{MLP}([E_{\text{item}}(x_i); f_i]) \in \mathbb{R}^d
    \end{equation}
    where $E_{\text{item}}$ denotes the item embedding table.
    
    \item \textbf{Context Augmentation}: Each objective embedding $\mathbf{O}_k$ is appended to the behavior sequence, and the combined representation subsequently serves as the input for the Causal Self-Attention mechanism.
   \begin{align}
        \mathbf{H}^{(0)} &= \{t_1, ..., t_n, \mathbf{O}_1, ...,\mathbf{O}_K\} \in \mathbb{R}^{d \times (n+K)}
    \end{align}
    where $\mathbf{t}_i$ denotes the $d$-dimensional embedding of the $i$-th user behavior.

    \item \textbf{Decoder Unit with Parameter-sharing design}: We adopt the architecture from \cite{touvron2023llama} with modifications of RMS normalization~\cite{rmsnorm} for gradient stabilization and casual masked self-attention for context-aware feature extraction across multiple objectives.
    Specifically, we share the query, key, value projection matrices across all $K$ objectives, reducing parameter complexity from $\mathcal{O}(K \cdot (n^2d+n d^2))$ to $\mathcal{O}(n^2d + n d^2)$ (see experiment ~\ref{subsub:para_share}).
        The attention computation follows:
        \begin{align}
        \mathbf Q^{(\ell)} &= \text{RMSNorm}(\mathbf H^{(\ell-1)})\mathbf W_{Q}^{(\ell)}, \\
        \mathbf K^{(\ell)} &= \text{RMSNorm}(\mathbf H^{(\ell-1)})\mathbf W_K^{(\ell)}, \\
        \mathbf V^{(\ell)} &= \text{RMSNorm}(\mathbf H^{(\ell-1)})\mathbf W_V^{(\ell)},\\
        \mathbf A^{(\ell)} &= \mathrm{Softmax}\!\left(\frac{\mathbf Q^{(\ell)}(\mathbf K^{(\ell)})^\top}{\sqrt{d_k}}\right),\\
        \mathbf H^{(\ell)} &= \mathrm{FFN}\!\bigl(\mathbf A^{(\ell)}\mathbf V^{(\ell)} + H^{(\ell-1)}\bigr).
        \end{align}
\end{enumerate} 

The user representation for the $k$-th objective is extracted from the output of the final layer ($L$) at the specific position corresponding to $\mathbf{O}_k$:
\begin{equation}
\mathbf{Emb}_u^k = \text{MLP}_k(\mathbf{H}^{(L)}[:, n + k])
\end{equation}

\subsubsection{Task-specific Personalized Decoder Unit}
To balance expressiveness and efficiency, we inject gated expert modules \emph{only} in the third decoder unit.  This single-layer design avoids the quadratic cost of stacking experts while still endowing the model with fine-grained, user-specific capacity.
We  enhance the standard feedforward network with gated expert modules:
\begin{equation}
    \text{FFN}_g(x) = \text{GeLU}(xW_{g1} + b_{g1})W_{g2} + b_{g2}
\end{equation}
\begin{equation}
    \text{FFN}_i(x) = \text{GeLU}(xW_{i1} + b_{i1})W_{i2} + b_{i2}
\end{equation}
\begin{equation}
    \text{gate\_score}_i = \text{softmax}(W_g[E_u; \text{FFN}_g(x)] + b_g)
\end{equation}
\begin{equation}
    \text{P-FFN}(x) = \sum_{i=1}^{\text{expertCut}} \text{gate\_score}_i \odot \text{FFN}_i(x)
\end{equation}
Let $\text{FFN}_g(x)$ denote the user representation generated by an MLP for the fixed positional feature $x$, where:
\begin{itemize}
    \item $E_u \in \mathbb{R}^d$ represents the user embedding vector obtained by concatenating user ID and device ID
    \item $\text{expertCut}(\cdot)$ implements top-$k$ expert selection to enhance computational efficiency
\end{itemize}

\subsection{Item Tower Architecture}

The item tower uses a \textbf{target-decoupled MLP ensemble} to generate task-specific item representations, reducing interference among competing objectives. This design decouples shared feature extraction from target-specific transformations, ensuring each task receives optimized representations.

To address optimization misalignment across diverse objectives (e.g., click-through rate, watch duration, conversion), we design target-specific transformation layers:
\begin{equation}
Emb_{I}^{(k)} = \text{MLP}_{k}([e_{\text{id}}, f_{\text{side}}]) \quad \forall k \in K
\end{equation}
Here, $e_{\text{id}}$ is the learnable embedding for the item's unique identifier, and $f_{\text{side}}$ includes auxiliary features like category tags, popularity metrics, and content-derived embeddings. Each $\text{MLP}_{k}$ is an independently parameterized multi-layer perceptron for task $k$, preventing gradient interference and enabling specialized feature adaptation for each objective.

\subsection{Multi-Objective Optimization}
\label{subsec:multi_objective_opt}

\subsubsection{Sampling Strategy}
We extend in-batch negatives: for every positive pair $(u,i)$, we keep three labels (pro\_lvr, max\_time, vtr) and draw $N-1$ shared negatives. This strategy ensures that each target focuses on its relevant positive samples while leveraging the shared negative samples across the entire batch, reducing the memory footprint from $\mathcal{O}(N^2)$ to $\mathcal{O}(N)$.

\subsubsection{Loss Formulation}
Our multi-task loss combines three objective specific components:
\begin{equation}
    \mathcal{L} = \sum_{k \in K} \alpha_k\mathcal{L}_k
\end{equation}
where K denotes the set of objectives.

Each objective loss follows a temperature-scaled softmax formulation:
\begin{equation}
    \mathcal{L}_k = -\frac{1}{N_k} \sum_{i=1}^{N} \log\left(\frac{\exp(\langle Emb_u^{k}, Emb_i^{k} \rangle/\tau)}{\sum_{j=1}^N \exp(\langle Emb_u^{k}, Emb_j^{k} \rangle/\tau)}\right)  \cdot l_{i,k}
\end{equation}

where $\langle \cdot,\cdot \rangle$ denotes inner product similarity, ${N}$ is the total number of samples in the batch, 
${N}_k$ is the number of positive samples for target $k$, $l_{i,k}$ is the binary label indicating whether sample $i$ is positive for target $k$, and $\tau$ is the temperature parameter.

The weights $\alpha_k$ are dynamically adjusted based on label distribution skewness:
\begin{equation}
    \alpha_k = \frac{\log(1+\gamma/|\mathcal{I}_k^+|)}{\sum_{k'} \log(1+\gamma/|\mathcal{I}_{k'}^+|)}
\end{equation}
where $\gamma$ controls the rebalancing intensity. This adaptive weighting mitigates the dominance of high-frequency objectives like clicks over rare conversions.

\subsection{Online Serving Architecture}
\label{subsec:serving}

Our multi-objective retrieval system employs a novel serving architecture that fundamentally differs from conventional single-objective deployments. The key components are:

\subsubsection{Multi-Objective Index Construction}
We maintain $K$ independent ANN indices $\{\mathcal{I}_1,...,\mathcal{I}_K\}$, where each index $\mathcal{I}_k$ contains item embeddings optimized for specific objective $k$:

\begin{equation}
    \mathcal{I}_k = \{\varphi_k(i) | i \in \mathcal{D}_{\text{pos}}^{(k)}\} \subset \mathbb{R}^{d}
\end{equation}

where $\mathcal{D}_{\text{pos}}^{(k)}$ denotes items with positive labels for objective $k$, and $\varphi_k: \mathcal{I} \rightarrow \mathbb{R}^d$ represents the objective-specific embedding function.

\subsubsection{Dynamic Quota Allocation}
To achieve personalized candidate distribution across objectives, we introduce adaptive quota allocation with three-stage optimization:

\begin{enumerate}
    \item \textbf{Task-Specific Weight Learning}: 
    For each user-item pair $(u,i)$, we learn objective-specific weights only over \emph{positive} instances—i.e., items that are ultimately exposed to the user and therefore recorded in our logging pipeline.
    \begin{equation}
        \mathcal{L}_{\text{quota}} = \frac{1}{|N|}\sum_{(u,i)\in N} \left(pscore_i - \sum_{k=1}^K w_{ik}s_k(u,i)\right)^2
    \end{equation}
    where $w_{ik} \in [0,1]$ denotes the adaptive weight for objective $k$, initialized as $1/K$, $pscore_{i}$ is the fused score of the fine-ranking stage, and $s_k(u,i)$ is the score for objective $k$ with stopped gradients.
    As negative items are never displayed, their $pscore$ values are undefined; restricting the loss to positives ensures reliable supervision without distribution-shift concerns.
    \item \textbf{User Preference Aggregation}: At serving time, aggregate weights from user's recent $n$ interactions:
    \begin{equation}
        \tilde{w}_u = \text{softmax}\left(\sum_{i\in\mathcal{H}_u} w_{i}\right) \in \mathbb{R}^K
    \end{equation}
    where $\mathcal{H}_u$ represents the user's last $n$ interacted items.
    
    \item \textbf{Adaptive Candidate Retrieval}: Allocate retrieval quotas proportionally:
    \begin{equation}
        Q_k(u) = \left\lfloor \tilde{w}_{uk} \cdot Q_{\text{total}} \right\rfloor
    \end{equation}
    with $Q_{\text{total}}$ being the total candidate budget per request.
\end{enumerate}

\subsubsection{Serving Pipeline}
The real-time serving workflow comprises three phases:

\begin{itemize}
    \item \textbf{Weight Calculation}: Compute $\tilde{w}_u$ via Algorithm 4.4.2
    \item \textbf{Parallel Retrieval}: Concurrent ANN searches on $\{\mathcal{I}_1,...,\mathcal{I}_K\}$ with $Q_k(u)$ quotas
    \item \textbf{Result Fusion}: Merge and deduplicate candidates across objectives
\end{itemize}

\section{Experiments}
\label{sec:experiments}
We rigorously evaluate the effectiveness and efficiency of our framework through both offline and online experiments on a large-scale industrial platform. The offline studies aim to validate its superiority over strong baselines and ablate key design choices, while the online A/B test confirms its substantial business impact in real-world deployment.

\subsection{Experimental Setup}
\label{subsec:setup}
\subsubsection{Datasets}
We evaluate our framework on Kuaishou's short-video recommendation platform -- one of the world's largest industrial systems serving over 400 million daily active users with 50 billion daily interactions. For offline evaluation, we construct a benchmark dataset using 6 consecutive days of production logs:

\begin{itemize}
    \item \textbf{Training Data}: 5 billion interactions from days 1-5
    \item \textbf{Test Data}: 1 billion interactions from day 6
    \item \textbf{Data Schema}: Each record contains \textit{(user id, item id, watch duration, click flag, timestamp, 15+ side features)}
\end{itemize}

Online evaluation is conducted through real-time A/B testing on live traffic spanning 7 days.

\setlength{\tabcolsep}{12pt} %
\begin{table*}[htbp]
\centering
\caption{\textbf{Offline Performance (\%) comparison with increase from second-highest to highest}}
\label{tab:performance}
\begin{tabular}{l|lcccc}
\toprule
\textbf{Category} & Metrics  & MVKE & ComiRec & KuaiFormer & \textbf{MPFormer}  \\
\midrule

\multirow{6}{*}{\centering\textbf{pro\_lvr}} 
& recall@10    & 3.72\%   & 18.56\%   & \underline{23.00\%}   & \textbf{33.30\%}  \\
& recall@50    & 15.44\%  & 40.46\%   & \underline{45.84\%}   & \textbf{56.55\%}  \\
& recall@100   & 27.57\%  & 52.99\%   & \underline{57.27\%}   & \textbf{66.15\%}  \\
& ndcg@1       & 0.54\%   & 4.41\%    & \underline{5.59\%}    & \textbf{9.90\%}  \\
& ndcg@10      & 1.83\%   & 10.46\%   & \underline{13.09\%}   & \textbf{20.27\%}  \\
& ndcg@100     & 6.24\%   & 17.26\%   & \underline{19.96\%}   & \textbf{26.98\%}  \\

\midrule

\multirow{6}{*}{\centering\textbf{max\_time}} 
& recall@10    & 3.09\%   & \underline{26.75\%}   & 16.77\%   & \textbf{41.26\%}  \\
& recall@50    & 11.79\%  & \underline{50.12\%}   & 36.99\%   & \textbf{63.91\%}  \\
& recall@100   & 19.92\%  & \underline{61.34\%}   & 48.87\%   & \textbf{71.89\%}  \\
& ndcg@1       & 0.39\%   & \underline{7.42\%}    & 3.76\%    & \textbf{13.59\%}  \\
& ndcg@10      & 1.47\%   & \underline{15.84\%}   & 9.31\%   & \textbf{26.03\%}  \\
& ndcg@100     & 4.63\%   & \underline{22.80\%}   & 15.63\%   & \textbf{32.38\%}  \\

\midrule

\multirow{6}{*}{\centering\textbf{vtr}} 
& recall@10    & 2.79\%   & \underline{21.58\%}   & 15.15\%   & \textbf{30.45\%}  \\
& recall@50    & 13.88\%  & \underline{44.09\%}   & 35.80\%   & \textbf{53.99\%}  \\
& recall@100   & 25.86\%  & \underline{55.23\%}   & 48.07\%   & \textbf{64.50\%}  \\
& ndcg@1       & 0.37\%   & \underline{5.35\%}    & 2.97\%    & \textbf{8.86\%}  \\
& ndcg@10      & 1.33\%   & \underline{12.34\%}   & 8.11\%   & \textbf{18.38\%}  \\
& ndcg@100     & 5.58\%   & \underline{19.08\%}   & 14.58\%   & \textbf{25.27\%}  \\

\bottomrule
\end{tabular}
\end{table*}

\subsubsection{Evaluation Metrics}
We adopt a dual evaluation protocol to assess system performance:

\begin{itemize}

    \item Recall@K: Proportion of relevant items in top-K predictions
    \item NDCG@K: Position-aware ranking quality metric
\end{itemize}

\subsubsection{Multiple Targets Setup } 
In our experiments, we define the following three objectives to capture diverse user preferences and evaluate the model's performance:

\begin{itemize}
    \item \textbf{Personalized Long-View Label(pro\_lvr)}: Positive samples are defined as items with viewing duration exceeding the 75th percentile in the user's recent 100-exposure sequence. This adaptive threshold captures individual viewing habits better than global criteria.
    
    \item \textbf{Surprise-Moment Label(max\_time)}: Samples with a viewing duration significantly longer than that of the past k items. 
    
    \item \textbf{VTR Label(vtr)}: The normalized target of the user's viewing duration.
\end{itemize}

Notably, these three objectives exhibit no strict inclusion relationship in their sample spaces. Quantitatively, the sample size follows a hierarchical pattern: the pro\_lvr objective generally yields more samples than the max\_time objective (15\% sample share), while the VTR objective (77\% sample share) bridges the two in scale. The pro\_lvr objective accounts for 83\% of the total samples, reflecting its broader data generation scope.

\begin{table*}[t]
    \centering
    \caption{Parameter Sharing Experiment:Task-independent qkv leads to a substantial increase in training and infer resources but only improves R@k by 1-2\%. Task-shared MLP in item tower leads to a 3\% - 6\% decrease in recall and a 6-8\% decrease in Ndcg.(R@k for Recall@k, N@k for Ndcg@k)}
    \label{tb:ablation}
    \resizebox{\textwidth}{!}{%
    \begin{tabular}{l*{12}{r}}
        \toprule
        \multirow{2}{*}{Category} & 
        \multicolumn{4}{c}{MPFormer} & 
        \multicolumn{4}{c}{MPFormer w item\_shared\_mlp} & 
        \multicolumn{4}{c}{MPFormer w independent\_QKV} \\
        \cmidrule(lr){2-5} \cmidrule(lr){6-9} \cmidrule(lr){10-13}
        & R@50 & R@100 & N@10 & N@100 & R@50 & R@100 & N@10 & N@100 & R@50 & R@100 & N@10 & N@100 \\
        \midrule
        pro\_lvtr  & 0.565 & 0.662 & 0.203 & 0.270 & 0.538 & 0.633 & 0.188 & 0.254  & 0.585 & 0.683 & 0.215 &  0.283\\
        max\_time  & 0.639 & 0.719 & 0.260 & 0.324 & 0.612 & 0.691 & 0.244 & 0.306  & 0.654 & 0.735 & 0.271 &  0.335\\
        vtr        & 0.540 & 0.645 & 0.184 & 0.253 & 0.528 & 0.626 & 0.183 & 0.249  & 0.556 & 0.662 & 0.192 &  0.262\\
        \bottomrule
    \end{tabular}%
    }
\end{table*}

\subsection{Offline Experimental Results}
\subsubsection{Comparison with Baselines}
\label{subsubsec:baselines}
We compare against three state-of-the-art industrial solutions:

\begin{itemize}
    \item \textbf{KuaiFormer} \cite{liu2024kuaiformer}: a Transformer-based, next-generation retrieval model with temporal attention.
    \item \textbf{ComiRec} \cite{cen2020controllable}: a model designed to capture and control multiple user interests for personalized recommendations. In our experiments, we employ four interest heads.
    \item \textbf{MVKE} \cite{xu2022mixture}: a model that combines multiple knowledge sources to improve recommendation accuracy. In our experiments, we introduce four virtual experts.
\end{itemize}

Table~\ref{tab:performance} presents a comprehensive comparison of offline performance among MPFormer and three baseline models: MVKE, ComiRec, and KuaiFormer, under three user behavior modeling objectives: pro\_lvr, max\_time, and vtr. The results clearly demonstrate the superior performance of MPFormer over existing state-of-the-art baselines. These substantial gains demonstrate the efficacy of our multi-task personalised sequential retriever in capturing fine-grained user intents at scale.

\begin{figure}[t]
    \centering
    \begin{subfigure}[t]{\linewidth}
        \includegraphics[width=\linewidth]{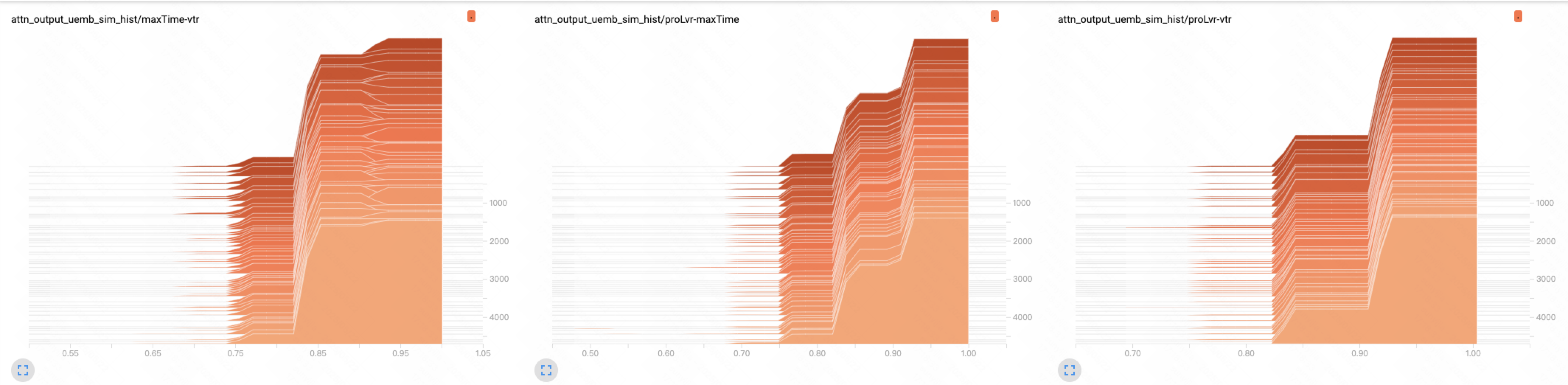}
        \caption{Inclusion of ID features: causes high similarity }
        \label{fig:uemb_hist_w_uid}
    \end{subfigure}
    \vspace{0.5em} %
    \begin{subfigure}[t]{\linewidth}
        \includegraphics[width=\linewidth]{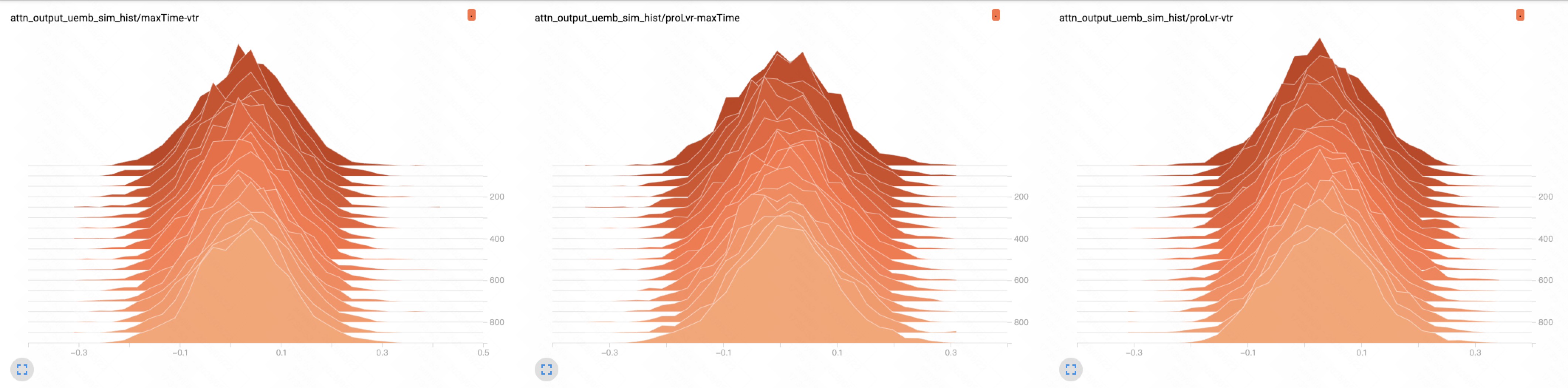}
        \caption{Exclusion of ID features: normally distributed near zero}
        \label{fig:uemb_hist_wo_uid}
    \end{subfigure}
    \caption{Impact of ID features on representation homogenization: (a) Inclusion of user/device IDs causes high cosine similarity between different objectives' user representations, indicating representation collapse into ID-dominated patterns; (b) Exclusion results in normally distributed similarity near zero, enabling objective-specific feature learning.}
    \label{fig:user_rep_comparison}
\end{figure}

\subsubsection{User Representation Analysis}
\label{subsec:user_rep}

We construct user representations using statistical attributes (e.g., age, location) and historical behavior sequences, intentionally avoiding user or device IDs. This choice is due to the homogenization effect caused by ID-based features.

Experiments show that including ID features increases the cosine similarity of user embeddings $U_k$ across objectives (Figure~\ref{fig:uemb_hist_w_uid}), indicating representation collapse. In contrast, removing ID features reduces embedding similarity (Figure~\ref{fig:uemb_hist_wo_uid}), enabling more discriminative representations per objective. 

The homogenization occurs because ID signals dominate learning, causing the model to rely on user-item co-occurrence rather than semantic feature interactions. As a result, representations degenerate into simplistic ID mappings, impairing generalization. Excluding ID features facilitates more nuanced, objective-specific representation learning, which is crucial for effective multi-objective recommendation.

\subsubsection{Parameter Sharing Mechanism Analysis}
\label{subsub:para_share}
1. \textbf{Shared QKV}.
In the Attention mechanism on the user side, we utilized shared QKV parameters across all objectives rather than assigning independent QKV for each target. This design choice was driven by empirical evidence from our experiments. Specifically, employing target-specific QKV parameters resulted in a 1.4× increase in model size and required 80\% more training machines. Despite these substantial resource requirements, this approach yielded only marginal improvements in Recall@100 (as shown in Table~\ref{tb:ablation}, approximately 2\% or less on average) compared to the shared baseline.

The computational complexity of the attention mechanism is significantly reduced by parameter sharing. This efficiency gain can be understood by considering the standard computational complexity of a self-attention Transformer block, which is generally expressed as $O(n^2 d + nd^2)$, where $n$ is the sequence length and $d$ is the embedding dimension. In our multi-objective scenario, this fundamental structure dictates the cost. For a model with batch size $B$, the complexity of using independent QKV parameters for each objective scales linearly with $K$, effectively becoming:
\begin{equation}
\label{eq:complexity_old}
\mathcal{O}_{\text{independent}} = K \cdot O\left((n+1)d^2 + (n+1)^2d\right)
\end{equation}
This is because the $O(n^2 d)$ and $O(n d^2)$ operations are repeated $K$ times.

In contrast, our shared QKV design integrates all $K$ target items into a single, unified attention computation. The effective sequence length for this operation becomes $(n + K)$, leading to a complexity of:
\begin{equation}
\label{eq:complexity_new}
\mathcal{O}{_\text{shared}} = O\left((n+K)d^2 + (n+K)^2d\right)
\end{equation}
This formulation captures the substantial efficiency gains achieved by sharing parameters across objectives, where the dominant quadratic terms no longer scale linearly with the number of objectives $K$.

2. \textbf{Task-independent MLP for Item Side.}
We conducted experiments comparing independent MLP and shared MLP architectures on the item side. The shared MLP approach aimed to generate a unified item embedding for all targets, relying solely on user-side representations to capture objective-specific semantics. However, as summarized in Table~\ref{tb:ablation}, this design led to significant degradation in offline metrics for each target (e.g., Recall@100 decreased by 5\%, Ndcg@100 decreased by 6\%).

This empirical evidence suggests that a single mapping function on the item side is inadequate to simultaneously satisfy the complex semantic and relational learning requirements of multiple targets. The inability to accurately capture target-specific item features stems from the intrinsic limitation of shared MLP structures in accommodating diverse learning objectives.

\subsection{Online Experiments}
\label{subsec:online_exp}
To evaluate the real-world impact of MPFormer, we conducted a 7-day large-scale A/B testing on Kuaishou’s short-video recommendation platform, involving randomly sampled daily active users (10\% of the total traffic). The experiment compares three configurations:
\begin{itemize}
    \item \textbf{Baseline}: Construct three independent retrieval pipelines for the targets \{pro\_lvr, max\_time, vtr\} using the Kuaiformer model, each returning $C/K = 1000$ candidates, with  a total capacity of $C=3000$.
    \item \textbf{MPFormer\_wo\_dq}: Construct three independent retrieval pipelines for the targets \{pro\_lvr, max\_time, vtr\} using the MPFormer model, each returning $C/K = 1000$ candidates (total capacity $C=3000$).
    \item \textbf{MPFormer\_w\_eq}: Within the unified multi-objective retrieval framework, a \emph{fixed and non-personalized} quota allocation is enforced for \emph{all} users according to Equation~\ref{eq:mean_quota}.
    \item \textbf{MPFormer}: Unified multi-objective retrieval with adaptive quota allocation (Section 4.4), maintaining the same total capacity (3000 items).
\end{itemize}

\begin{table}[h]
\centering
\caption{Online A/B Test Results (Relative Improvement)}
\label{tab:online}
\begin{tabular}{l@{\hskip 6pt}c@{\hskip 6pt}c@{\hskip 6pt}c}
\toprule
 & \_wo\_dq & \_w\_eq  & \textbf{MPFormer} \\
\midrule
\makecell{Total Watch Time} & +0.402\% & +0.420\% & \textbf{+0.426\%} \\
\makecell{Total App Usage Time}  & +0.165\% & +0.143\% & \textbf{+0.195\%} \\
\makecell{Average View Duration} & +0.866\% & +0.613\%  & \textbf{+0.455\%} \\
\makecell{Real Show}  & -0.840\% & -0.618\% & \textbf{-0.411\%}\\
\bottomrule
\end{tabular}
\end{table}

\subsubsection{Performance Metrics}
Table~\ref{tab:online} summarizes the key improvements in engagement metrics:
\begin{itemize}
    \item \textbf{Metrics}: We report Total Watch Time and Total App Usage Time to evaluate user engagement in terms of overall platform usage, while Average View Duration measures the mean time spent per video during recommended sessions.
    \item \textbf{Online A/B Test Results}: Online A/B testing results demonstrate that MPFormer achieves a +0.426\% increase in total watch time, a +0.195\% increase in total app usage time, and a +0.455\% improvement in average view duration.
\end{itemize}

These results indicate that MPFormer delivers significantly better recommendation outcomes and contributes to substantial revenue gains for the platform.

\subsubsection{System Efficiency}
We evaluate the system efficiency of MPFormer (a unified model trained for $K$ objectives and inferenced in one pass) against the baseline of $K$ separately trained and serviced models. MPFormer demonstrates superior efficiency across all aspects:
\begin{itemize}
    \item \textbf{Training Cost}: Consolidating $K$ model trainings into single-model paradigm increases resource consumption by 13\% compared to single-objective training, but achieves 60\% reduction relative to separate training of $K$ models.
    \item \textbf{Inference Cost}: The unified architecture generates embeddings for all three objectives through single forward pass, reducing serving resource requirements by 66.7\% under equivalent QPS compared to running $K$ separate models.
    \item \textbf{Latency}: Maintains stable P99 latency at 80 ms, showing no significant increase compared to single-objective models.
\end{itemize}

\begin{figure}[t]
\centering
\includegraphics[width=\linewidth]{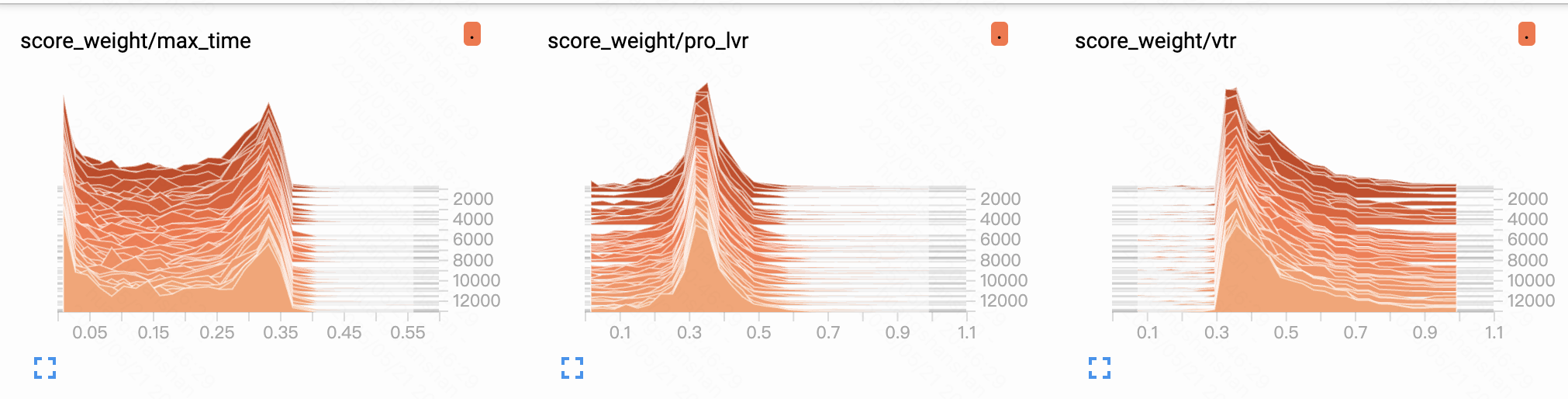}
\caption{Analysis of Dynamic Quota: item-side weight distributions across three objectives (max\_time, pro\_lvr, vtr) demonstrate distinct patterns, with the Y-axis representing the proportion of weights: max\_time weights are widely dispersed, pro\_lvr peaks near $1/K$, while vtr concentrates above $1/K$, reflecting their different optimization characteristics.}
\label{fig:photo_hist}
\end{figure}

\subsubsection{Dynamic Quota Analysis}
Our analysis of the dynamic quota mechanism reveals three key insights:

\begin{itemize}
\item \textbf{Photo-side Weight Distribution}: Figure~\ref{fig:photo_hist} demonstrates distinct patterns across objectives:
\begin{itemize}
    \item \textit{max\_time} weights exhibit high dispersion (mean $<$ 0.33, SD=0.21) due to objective sparsity
    \item \textit{pro\_lvr} follows a normal-like distribution peaked near $1/K$ (mean=0.33, $\sigma$=0.09)
    \item \textit{vtr} shows positive skewness (median=0.41) as it aligns with the core optimization metric of watch duration
\end{itemize}

\item \textbf{User-side Quota Allocation}: Normalized weights per user show systematic variation:
\begin{equation}
\label{eq:mean_quota}
    \text{mean quota allocation} = 
    \begin{cases}
        35\% & \text{pro\_lvr} \\
        18\% & \text{max\_time} \\
        47\% & \text{vtr}
    \end{cases}
\end{equation}
The percentage point divergence from photo-side distributions confirms the necessity of user-level adaptation.

\item \textbf{System Impact}: Dynamic quota allocation significantly improves the exposure rate across the multi-object recall framework, resulting in a 12.7\% increase (from 6.76\% to 7.62\%). Furthermore, it contributes to enhanced posterior efficiency, as evidenced by a 1.3\% uplift in effective-view rate. Specifically, as shown in Table~\ref{tab:online}, it contributes a 0.03\% gain in total duration and effectively mitigates a 0.43\% drop in realshow, which denotes the absolute number of videos actually displayed to users after passing all ranking stages. Compared with the fixed-ratio variant MPFormer\_w\_eq, MPFormer further curbs the real-show loss (–0.411 \% vs. –0.618 \%), confirming the incremental benefit of user-level adaptation.

\end{itemize}

\section{Conclusion}
\label{sec:conclusion}
In this work, we address the critical challenge of multi-stage optimization misalignment in industrial recommendation systems, where the semantic gap between multi-objective ranking paradigms and single-objective retrieval modeling limits system efficiency and scalability. To bridge this gap, we propose MPFormer, a dynamic multi-task Transformer framework that introduces three key innovations: (1) an objective-conditioned transformer for joint encoding of user behavior and multi-task semantics, (2) personalized target weights enabling dynamic retrieval adjustment, and (3) deep integration of user personalization into token representations and model architecture. Deployed in Kuaishou’s short-video recommendation system, MPFormer demonstrates significant practical value by stably serving over 400 million daily active users while boosting user engagement and operational efficiency. Compared to traditional parallel multi-path solutions, our framework eliminates the linear resource overhead and offers superior flexibility in handling loosely coupled objectives. This work not only provides an efficient and scalable paradigm for multi-objective retrieval but also advances the industrial applicability of dynamic multi-task learning, paving the way for future research on adaptive recommendation systems in large-scale real-world scenarios.

\bibliographystyle{ACM-Reference-Format}
\bibliography{sample-base}

\typeout{get arXiv to do 4 passes: Label(s) may have changed. Rerun}

\end{document}

%% file: model_arch1.tex
\begin{figure*} %
    \centering
    \includegraphics[width=0.9\textwidth]{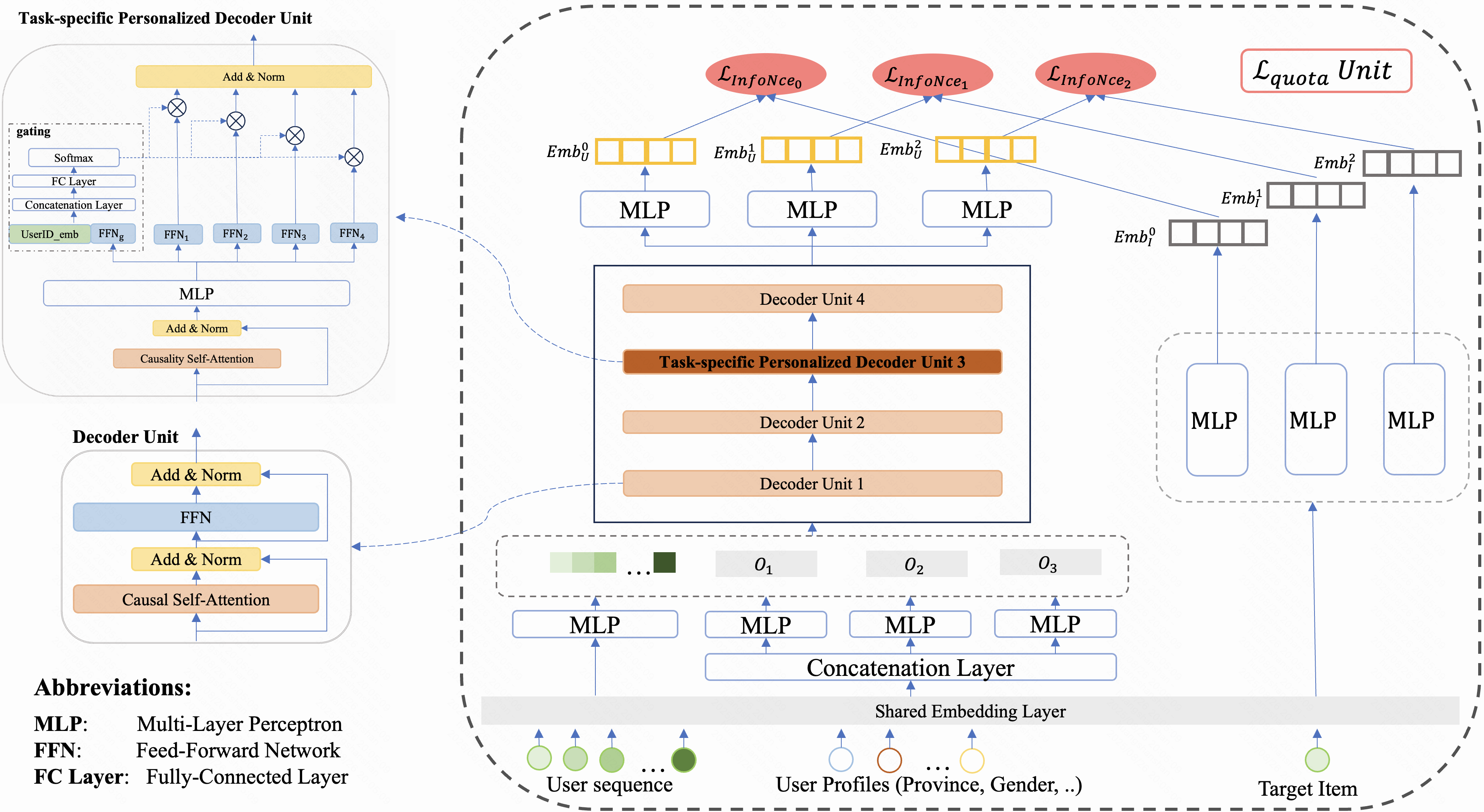} %
    \caption{ Architecture of the Proposed MPFormer Model. (a) User-side inputs consist of two components: task-specific personalized tokens generated by applying $K$ independent MLPs to user profile features, and recent watch history sequences. (b) A four-layer transformer stack processes the inputs; the third layer is instantiated as our \textbf{PFormer} to inject fine-grained user personalization, while the remaining layers follow the standard \textbf{Decoder Unit} design. The final $K$ hidden states are extracted as multi-task representations. (c) The joint loss function combines InfoNCE losses for all $K$ tasks with the dynamic quota allocation loss ($\mathcal{L}_{quota}$), where the latter's mechanism is detailed in Figure~\ref{fig:fig_model2}.}
    \label{fig:fig_model1}
\end{figure*}

%% file: quota_fig.tex
\begin{figure*} %
    \centering
    \includegraphics[width=0.9\textwidth]{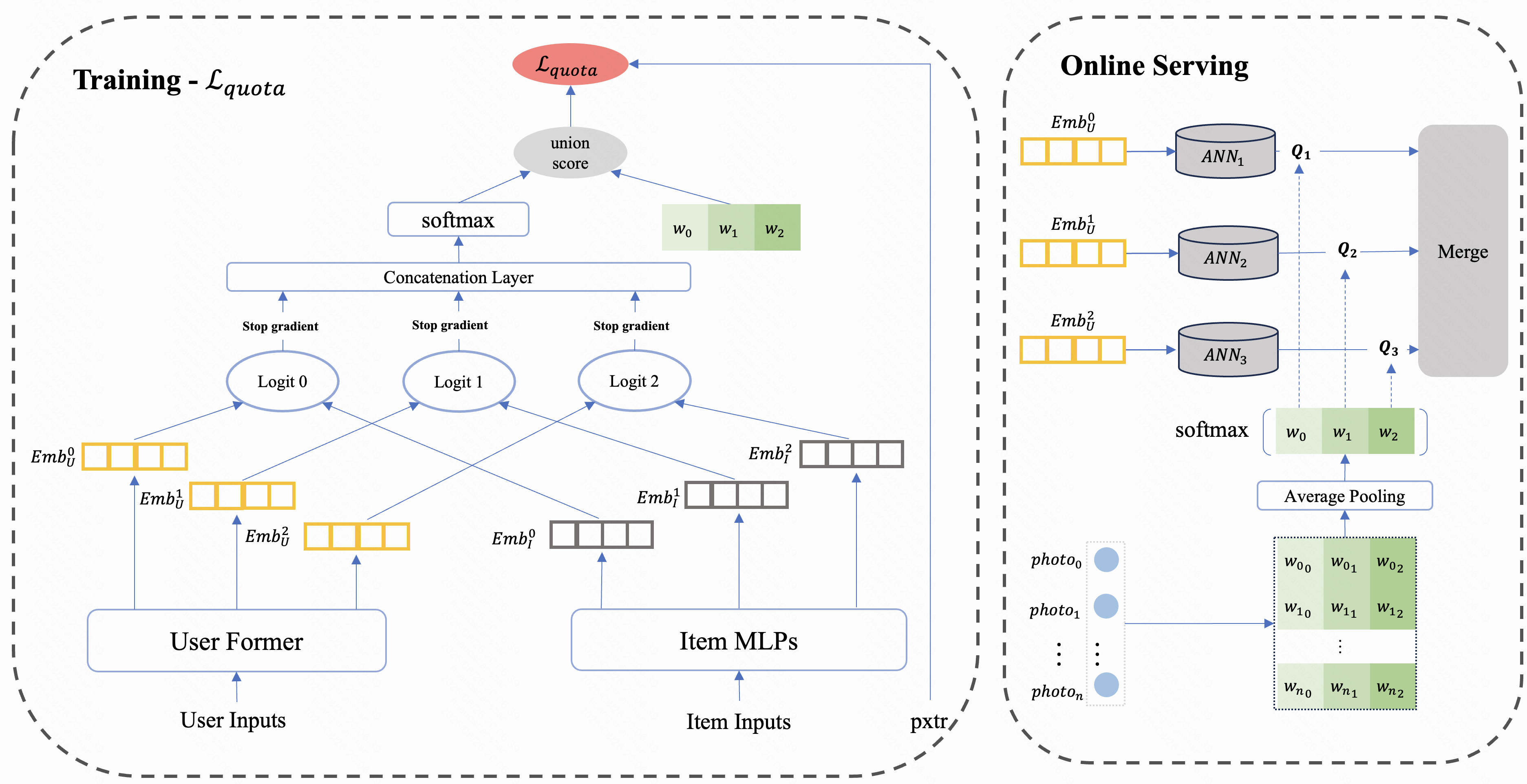} %
    \caption{ Dynamic Quota Allocation Mechanism. (a) Loss Computation($\mathcal{L}_{quota}$): For each item $i$ and objective $k$, the weight $w_{ik}$ is learned using the ranking scores from the subsequent stage as supervision signals. 
(b) Online Serving: During inference, the system computes per-objective weights $\{w_k\}_{k=1}^K$ by aggregating historical interactions over $n$ recent items. These weights $\{w_k\}_{k=1}^K$ are obtained from the training phase by optimizing $\mathcal{L}_{quota}$ for each target item. The system then dynamically adjusts the retrieval quota allocation based on these weights. }
    \label{fig:fig_model2}
\end{figure*}